\documentstyle[multicol,prb,graphics,aps]{revtex}
\begin{document}

\title{The mean free path for electron conduction in metallic fullerenes}

\author{O. Gunnarsson and J.E. Han\\   
Max-Planck-Institut f\"ur Festk\"orperforschung, 
D-70506 Stuttgart, Germany}

\maketitle

\begin{multicols}{2}
The electrical resistivity $\rho$ of a metal is usually interpreted 
in terms of the average distance $l$ an electron travels before it is 
scattered. As the temperature $T$ is raised, $\rho$ is increased and 
the apparent mean free path $l$ is correspondingly reduced. In this 
semi-classical picture $l$ cannot be (much) shorter than the distance 
$d$ between two atoms.  This has been confirmed for many systems and 
was considered a universal behaviour.\cite{Fisk,Allen} Recently, some
apparent exceptions were found, e.g., alkali-doped 
Fullerenes\cite{Hebard,Degiorgi,Hou,Housat,RMP} and  high temperature
superconductors. These systems could, however, be in exotic states 
where only a small fraction of the conduction electrons contribute 
to the conductivity, and $l$ would then have to be correspondingly 
larger to explain the observed $\rho$.  
Here we report on an essentially exact calculation for a model of 
alkali-doped Fullerenes, where the electrons are scattered by 
intramolecular vibrations. The resistivity at large $T$ 
corresponds to $l\ll d$, demonstrating that there is no fundamental 
principle of the type $l >d$. At large $T$ the semi-classical picture 
breaks down and the electrons cannot be described as quasi-particles. 
 
We have also calculated the resistivity due to {\it electron-electron} 
scattering for a half-filled Hubbard model. In this case the 
resistivity saturates and $l$ is not very much smaller than $d$. This
difference is traced to the difference between bosons and fermions.  
The resistivity is often calculated using the Boltzmann equation.
Although this equation is usually derived semi-classically, assuming 
$l \gg d$, in our model for electron-vibration scattering it does not 
break down qualitatively at large $T$, where $l \ll d$. For small 
$T$ the calculated $\rho$ due to electron-vibration scattering has 
a linear dependence on $T$ and a strong dependence on the pressure, 
in agreement with experiment.\cite{Vareka} 

For A$_3$C$_{60}$ (A= K, Rb) at $T\sim 500$ K, the resistivity 
is $\rho \sim 2-5$ m$\Omega$cm. Assuming a spherical Fermi surface 
and three conduction electrons per site,\cite{Hebard} we find that 
$l\sim 1-2$ \AA \ is almost an order of magnitude smaller than the 
separation $d=10$ \AA \ of the C$_{60}$ molecules. The 
experimental data have substantial uncertainties, but this 
is unlikely to influence the qualitative discussion of $l$.              
Different experimental methods (direct and optical) for
different types of samples (thin  films and doped single crystals)
all suggest that $l \ll d$ for large $T$. It has been suggested 
that the large $\rho$ is due to the electrons being scattered 
inside the molecules. This would imply scattering between bands 
separated by at least 1 eV, and it plays a fairly small role for 
experimental $T$. In addition, we find that this interband 
scattering\cite{Allen} {\it reduces} the resistivity by providing 
an additional channel for conduction.

The conduction in 
A$_3$C$_{60}$ takes place in a partly filled $t_{1u}$ band. 
The $T$-dependent part of the resistivity is assumed to be due 
to scattering against phonons (vibrations) with H$_g$ symmetry. 
We therefore consider a model with a three-fold degenerate 
$t_{1u}$ level  and a five-fold degenerate H$_g$ Jahn-Teller phonon
on each molecule, the hopping between the molecules and the 
coupling between the electrons and the phonons. The hopping 
takes into account\cite{disorder,ising} the orientational 
disorder of the molecules.\cite{Stephens} The one-particle 
band width is $W=0.6$ eV and the phonon frequency is 
$\omega_{ph}$. The coupling is determined by the dimensionless 
$\lambda=(5/3)N(0)g^2/\omega_{ph}$, where $N(0)$ is the density 
of states per spin, $g$ is the electron-phonon coupling constant 
and $\omega_{ph}$ is the phonon frequency.

We perform a finite temperature QMC calculation,\cite{Scalapino} 
treating the phonons quantum mechanically.  We calculate the
current-current correlation function for imaginary times and
make a transformation to real frequencies, using a maximum
entropy method.\cite{Jarrell} The QMC method has no ``sign-problem'',
and the resistivity of the model can be calculated 
essentially exactly down to quite small $T$. 

\begin{figure}[bt]
\centerline{
\rotatebox{270}{\resizebox{!}{3in}{\includegraphics{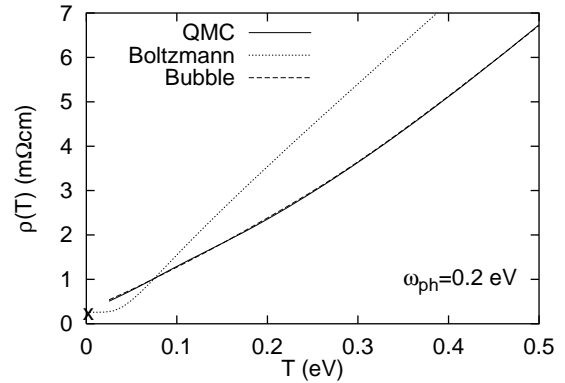}}}}
\caption[]{\label{fig:res}The resistivity as a function of $T$.        
Results of the full QMC calculation, the Boltzmann equation 
(Bloch-Gr\"uneisen) and the bubble diagram are shown. The symbol $\times$
shows the $\rho(T=0)$ due to the orientational disorder. The figure 
illustrates that $\rho$ can become extremely large, that the bubble 
calculation is quite accurate and that there is no qualitative 
break-down of the Boltzmann equation at high $T$.}
\end{figure}

Fig. \ref{fig:res} shows the resistivity for a cluster of 48 
C$_{60}$ molecules for $\lambda=0.5$ and $\omega_{ph}=0.2$ eV. 
The QMC calculation (full line) shows that the resistivity can 
become very large, corresponding to $l\sim 0.7$ \AA  \ at $T=0.5$ 
eV. By also considering an unrealisticly large $T$, we emphasise 
the lack of a limitation of the type $l>d$. Then there cannot be 
any general principle requiring $l>d$, because such a principle
would then also apply to the present model. Qualitatively similar 
result were obtained by Millis {\it et al.},\cite{Millis} using 
the dynamical mean-field theory (DMFT)\cite{DMF} and assuming 
classical phonons. Their calculation, however, does not prove 
that $l\ll d$ is possible, since it involves approximations. 
The moderate differences to our results are probably due to 
different models and their use of classical phonons.  

\begin{figure}[bt]
\centerline{ 
\rotatebox{0}{\resizebox{2in}{!}{\includegraphics{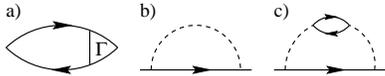}}}}
\vskip0.2cm
\caption[]{\label{fig:diag}
Relevant diagrams.  The current-current response function (a) and  
two approximations to the electron self-energy ((b) and (c)) are shown. 
The full and dashed lines represent electron and phonon Green's functions.  
Self-consistent Green's functions are used in (a) but not in (b) or  (c).
}
\end{figure}

While the QMC calculation is essentially exact, it is 
hard to interpret the results. For this purpose 
we use a diagrammatic approach. The idea is to simplify the 
diagrams as far as possible without loosing the qualitative 
agreement with the QMC results. These diagrams are then studied 
to obtain a qualitative understanding of the results. This
approach (Kubo formalism) requires the calculation 
of a bubble diagram including vertex corrections (see Fig. 
\ref{fig:diag}a).  We neglect the vertex and calculate the bubble 
using the electron Green's function from the QMC calculation.  
The resulting resistivity (dashed line  in Fig. \ref{fig:res}) 
is practically identical to the QMC result, 
justifying the neglect of vertex corrections for the present model.    
It was shown by Holstein\cite{Holstein} that in the limit of a broad
electronic band, all vertex corrections except ladder diagrams can be 
neglected and that a Boltzmann equation can be derived. Holstein's      
derivation is not valid for the narrow band considered here, but our  
calculations show that his arguments are still rather accurate. 
For our  model with a 
${\bf q}$-independent electron-phonon coupling, even the ladder 
diagrams can be neglected. Essentially following Holstein
we obtain approximately a Boltzmann like conductivity
\begin{equation}\label{eq:2}
\sigma(T)\sim \int d\omega N(\omega) (-{df(\omega)\over d\omega})
{1\over {\rm Im} \Sigma(\omega)}|j_k|^2_{\varepsilon_k=\omega},
\end{equation}
where $N(\omega)$ is the density of states, $f$ is the
Fermi function, $\Sigma(\omega)$ is the electron self-energy,
$j_k$ is the current matrix element for a state with the label $k$ and 
the energy $\varepsilon_k$. We interpret Im $\Sigma$ as the inverse
of the relaxation time. For a large $T$, Im $\Sigma$ becomes 
comparable to or larger than the one-particle band width. Since Im $\Sigma$,
is related to the inverse life-time of the quasi-particles,
this means that the life-times become so short that the concept
of a quasi-particle breaks down at large $T$.

The resistivity thus depends crucially on $\Sigma$.     
To understand its behavior, we consider the diagram
in Fig. \ref{fig:diag}b calculated with bare Green's functions
and for simplicity neglecting the orbital degeneracy
\begin{eqnarray}\label{eq:3}
&&\Sigma^{(1)}({\bf k},\omega) \\
&&=g^2\sum_{\bf q}\lbrack
{n_B(\omega_{ph}) +1 -f(\varepsilon_{\bf q} )\over \omega-\omega_{ph}-
\varepsilon_{\bf q} } 
+{n_B(\omega_{ph}) +f(\varepsilon_{\bf q} )\over \omega+\omega_{ph}-
\varepsilon_{\bf q} }\rbrack, \nonumber
\end{eqnarray}
where 
\begin{equation}\label{eq:4}
n_B(\omega_{ph})={1\over e^{\omega_{ph} /T}-1} \ \ {{ \longrightarrow} 
\atop T\to \infty} \ \ \ {T\over \omega_{ph}}
\end{equation}
is the Bose  occupation number. For large $T$,  $n_B$ becomes 
large, leading to a large Im $\Sigma$, a small $\sigma$ 
and a large $\rho$. The Bose nature of the phonons is therefore
of crucial importance for our result.

\begin{figure}[bt]
\centerline{ 
\rotatebox{-90}{\resizebox{!}{2.5in}{\includegraphics{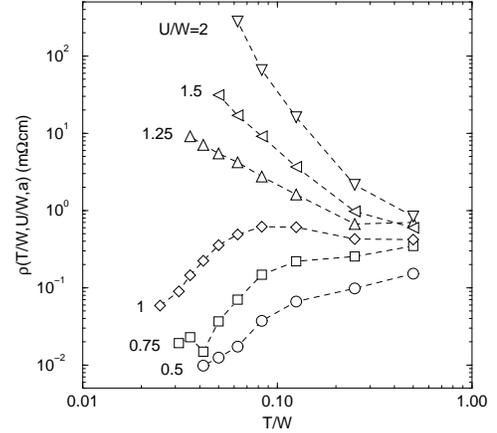}}}}
\caption[]{\label{fig:elelres}
The resistivity of the nondegenerate Hubbard model.
Results are shown for different values of the ratio between the
Coulomb repulsion $U$ and the band width $W$. The figure illustrates 
how the resistivity saturates for large $T$ in the electron-electron 
scattering model.}             
\end{figure}
This is further illustrated by comparing with the resistivity 
due to the electron-electron scattering. Using the DMFT 
we have calculated the resistivity for a nondegenerate Hubbard 
model on a Bethe lattice with $d=1$ \AA. We focus on the 
half-filled case, which is relevant for A$_3$C$_{60}$, and we do 
not consider the case of a doped Mott insulator.\cite{linearT}
The impurity problem is solved with a QMC method. Fig. 
\ref{fig:elelres} 
shows $\rho(T)$ for different values of the on-site Coulomb 
interaction $U$. For $U/W<1$ the system is a metal and $\rho(T)$ grows
with $T$, while for $U/W>1$ it is an insulator and $\rho(T)$ decreases
with $T$. The important observation is that in the metallic case 
$\rho(T)$ saturates at $\rho\sim 0.4$ m$\Omega$cm, which corresponds 
to $l/d\sim 1/3$. Thus, in contrast to the electron-phonon scattering 
case,  $l$ is not very much smaller than $d$ in the metallic state of 
this model and within DMFT.                     

To understand this, we study the electron self-energy $\Sigma$ to 
second order in $U$, since $\Sigma$ determines $\rho$ in the DMFT.  
For low $T$, there is little scattering due to the small phase 
space available, as controlled by the Fermi functions. As $T$ 
increases, the available phase space grows and $\rho$ increases. 
However, for large $T$,  $\rho$ essentially saturates, since the 
Fermi functions approach a constant value. This is in strong contrast 
to the Bose occupation numbers (Eq.~(\ref{eq:4})), which keep 
increasing with $T$. The qualitative  difference between the two 
models for large $T$ can then be traced to the 
difference between fermions and bosons.

We now address the validity of the Boltzmann equation for
the case of the electron-phonon scattering in view of $l\ll d$.
We have calculated the resistivity using the Ziman version 
of the Bloch-Gr\"uneisen solution of the Boltzmann 
equation\cite{Grimvall} and added the resistivity due to the 
orientational disorder as a $T$-independent contribution.\cite{Gelfand}
Using the plasma frequency $\omega_p=1.2$ eV, we obtain the 
dotted line in Fig. \ref{fig:res}. The Boltzmann result is larger 
than the QMC result for large $T$, but there is no qualitative 
break down of the Boltzmann equation, although $l\ll d$ and the 
quasi-particle concept is not applicable. The justification for 
the Boltzmann equation in this  limit is not the semi-classical 
derivation, but the (approximate) derivation  from the full 
quantum mechanical Kubo formulation (Eq.~(\ref{eq:2})). The 
proper language is not in terms of a very short mean free path,  
but in terms of a very broad spectral function (Im $\Sigma$ large),
as discussed above.

The QMC calculation gives an approximately linear $T$ dependence.
This agrees with the experimental result that $\rho$ is linear 
down to about 100-200 K.\cite{Vareka} although the linearity is 
not as pronounced as for some high temperature superconductors. 
The result may seem surprising, since at 
small $T$ the probability of exciting finite energy phonons is 
exponentially small as is the contribution to $\rho$. Calculating 
the bubble diagram in Fig. \ref{fig:diag}a using electron Green's 
functions with the self-energy in Fig. \ref{fig:diag}b indeed gives an 
exponential behaviour. The use of the QMC Green's function, however,  
gives a linear behavior, in agreement with the full QMC calculation.
The reason is that the QMC Green's function also involves processes 
like in Fig. \ref{fig:diag}c, where a virtual phonon is created
 followed by the decay of this phonon into an electron-hole 
pair.\cite{quadratic} The excitation energy of such a pair can be 
arbitrarily small, leading to a quadratic $T$-dependence for 
$\rho$.\cite{quadratic} In our model this  goes over in an approximately 
linear behavior already for very small $T$.

Calculations for different values of $\lambda$ show a strong 
$\lambda$-dependence for small $T$. This is in agreement with
the strong pressure dependence observed experimentally,\cite{Vareka}
since the application of pressure increases the band width 
and reduces $N(0)$ and thereby $\lambda$. For $\lambda \sim 1$
there is a transition to an insulating state. 

Our use of a fairly realistic model with a three-fold degenerate
electronic level and a five-fold degenerate phonon is     
not crucial for the qualitative behaviour of $\rho(T)$, like
$l \ll d$ for large $T$. An essential feature is,  however, the
use of intramolecular phonons acting on a narrow band.
The phonon induced shifts ($\sim T$) of the $t_{1u}$ levels  
can be substantial compared with the small band width. It would 
be interesting to perform calculations for a model where the phonons 
couple to the hopping matrix elements and where interband transitions
become important for large $T$. Such a model would be more relevant 
for, e.g., transition metal compounds, where the resistivity 
typically saturates when $l \sim d$.

{\bf Acknowledgements}

We are grateful to M. Jarrell for making his MaxEnt program available
and E. Koch for a careful reading of the manuscript.
The work has been supported by the Max-Planck-Forschungspreis.

Correspondence and requests for materials should addressed to O.G.

(e-mail gunnar@and.mpi-stuttgart.mpg.de)

\end{multicols}

\end{document}